# Energy efficient manipulation of topologically protected states in non-volatile ultrafast charge configuration memory devices

Anže Mraz*[1,4], Rok Venturini[1,3], Michele Diego[1], Andrej Kranjec[1], Damjan Svetin[1,2], Yaroslav Gerasimenko[1,2], Vitomir Sever[1], Ian A. Mihailovic[1], Jan Ravnik[1,5], Igor Vaskivskyi[1,2], Maria D'Antuono[6], Daniela Stornaiulo[6], Francesco Tafuri[6], Dimitrios Kazazis[5], Yasin Ekinci[5] and Dragan Mihailovic*[1,2]

[1] *Jozef Stefan Institute, Dept. of Complex Matter, Jamova 39, SI-1000 Ljubljana, Slovenia*

[2] *CENN Nanocenter, Jamova 39, SI-1000 Ljubljana, Sloveniam*

[3] *Faculty for Mathematics and Physics, University of Ljubljana, Jadranska 19, SI-1000 Ljubljana, Slovenia*

[4] *Faculty for Electrical Engineering, University of Ljubljana, Tržaška 25, SI-1000 Ljubljana, Slovenia*

[5] *LMN-Paul Scherrer institute, Villigen, Switzerland*

[6] *Dept. of Physics, University of Naples and CNR-SPIN, Naples, Italy*

**Non-volatile magnetic storage, from 1940s magnetic core to present day racetrack memory[1] and magnetic anisotropy switching devices[2–4] rely on the metastability of magnetic domains to store information. However, the inherent inefficiency of converting the information-carrying charge current into magnetization switching sets fundamental limitations in energy consumption. Other non-magnetic non-volatile memories such as memristors, ferroelectric memory[5] and phase change memory devices also rely on energetically relatively costly crystal structural rearrangements to store information[6,7]. In contrast, conventional electronic charge states in quantum dots for example, can be switched in femtoseconds with high efficiency, but any stored information dissipates rapidly[8]. Here we present a radically different approach in the form of a charge-configuration memory (CCM) device that relies on charge-injection-driven electronic crystal melting and topological protection of the resulting electronic domain configurations of a two-dimensional electronic crystal to store information. With multiprobe scanning tunneling microscopy (STM) we show microscopically, within an operational device, how dislocations in the domain ordering lead to metastability by a mechanism that is topologically equivalent to magnetic bubble memory. The devices have a very small switching energy (< 2.2 fJ/bit), ultrafast switching speed of <11 ps and operational range over more than 3 orders of magnitude in temperature (< 250 mK ~ 190 K). Together with their simple functionality, a large resistance**

**switching ratio, straightforward fabrication and impressive endurance, CCM devices introduce a new memory paradigm in emerging cryo-computing[9] and other high-performance computing applications that require ultrahigh speed and low energy consumption.**

Manipulation of charge, rather than spin, would potentially increase the performance of memory devices, avoiding the weak coupling between charge and spin degrees of freedom that limits the efficiency of magnetic memories. However, the problem with charge ordered states in general is that they are strongly coupled to charge fluctuations of the environment and quickly dissipate. The desirable metastability that comes with orientation of ferromagnetic or ferroelectric domains is not trivial to achieve in electronic states. One possible solution is utilized in ferroelectric memories where the ferroelectric polarization is switched directly by the electric field and the polarization of ionic positions within domains defines the non-volatile information state[5]. Electronic avalanche breakdown in narrow gap Mott insulators is also a possibility[10]. Unfortunately these are still inherently energetically costly and rather slow[5,10]. Another possibility, which is explored here is to use domains in an electronic crystal to store information whereby the relative position of electrons on the lattice is manipulated by charge injection. Instead of switching a vector or pseudovector quantity as is the case of a ferroelectric polarization or magnetization respectively, here reconfiguration of charges within domains relative to each other is used to store information. Addressing the problem of dissipation, we show that stability of such charge domain ordering can be enhanced by topologically protected defects, which impart stability to the different domain configurational states. The problem of how to read and write the information stored within domain configurations is solved in the present case by the fact that different domain configurations have different electrical resistance. Having achieved non-volatility and reproducibility, the result is an exceptionally simple memory device with outstanding performance.

The device relies on the fact that charge-injection causes frustrated compression of electronic crystal order in the layered transition metallic dichalcogenide 1T-TaS$_2$. In common with many similar materials, at low temperatures 1T-TaS$_2$ exhibits a uniform commensurate (C) electronic order with a periodicity that is a multiple of the underlying lattice constants[11]. The electrons, shown in the atomic resolution STM image (Fig. 1a) are subject to Coulomb interaction and are packed into a regular hexagonal 'polaron crystal' lattice[12] (Fig. 1d), with one electron localized on every 13$^{th}$ Ta metal site[13]. The localized electrons cause a polaronic distortion of the surrounding lattice in the form of a star of David (Fig. 1b). The unusual feature of this material is the 0.3 eV band gap in the C state[14,15] which is commonly attributed to a Mott state within TaS$_2$ layers that in combination with a pairwise charge stacking of polarons[16] perpendicular to the layers determines the electronic transport properties[17]. Chemical doping of such a state introduces weakly pinned dopant-induced dislocation defects that destroy long-range translational order, while retaining orientational order[18]. However, recent work shows that chemical doping is not essential for the appearance of such hexatic order; domain walls can appear because they can accommodate extra electrons[12,19]. This feature cannot be described by traditional charge-density wave physics such as doping of Fermi surface nested systems, and cannot be predicted by band structure calculations[20,21], but can be readily understood either microscopically in terms of electron correlations[12,19], or phenomenologically in terms of Landau theory[22].

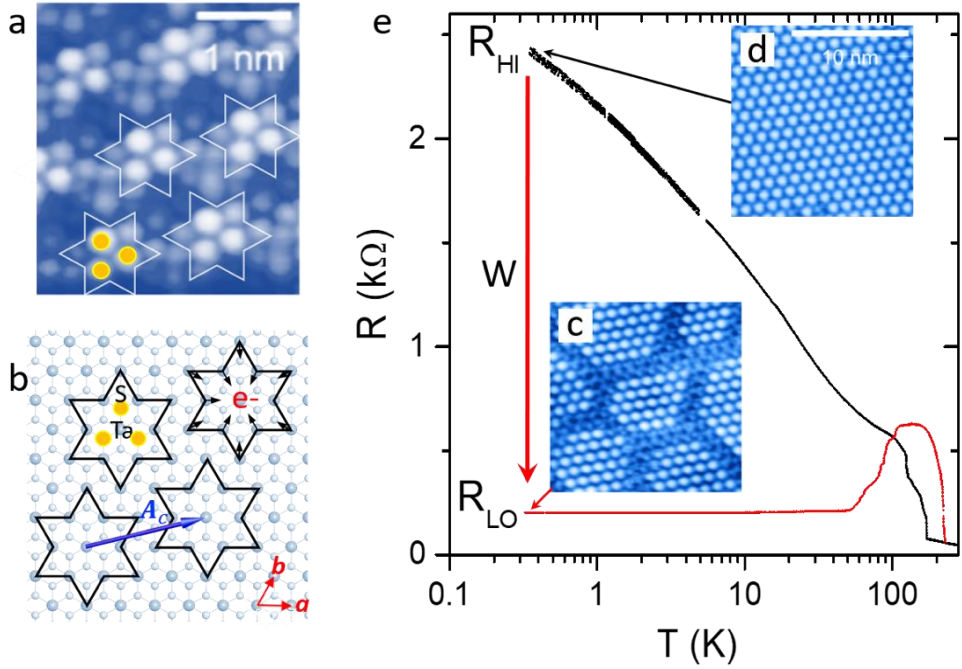

**Figure 1. Charge order configurations and resistance switching in 1T-TaS$_2$.** The electronic charge configurations associated with switching are shown by 4K STM images: a) atomic resolution image of the C charge order with enhanced S orbitals (yellow). b) A schematic of the C structure on the 1T-TaS$_2$ electronic lattice. Ta atoms are displaced towards the extra electron on the center Ta (shown) forming a star. The crystal ***a*** and ***b*** axes (red) and electronic lattice vector ***A**$_c$* (blue) are indicated. c) An STM image of the metallic metastable domain structure induced by charge injection on a mesoscopic scale. d) An STM image of the insulating C charge order on a mesoscopic scale. e) The temperature dependence of the 4-contact resistance upon switching. In this case an electrical W pulse causes the transition from $R_{HI}$ to $R_{LO}$ at 250 mK as indicated by the arrow.

The C state in 1T-TaS$_2$ uniquely responds to external perturbations in a way that makes it useful for a non-volatile memory device. Photoexcitation[23] or direct charge injection[24–27] breaks apart the long range charge order within < 1 ps[28], creating a complex domain structure without the presence of chemical-doping-induced defects, as shown in the STM image in Fig. 1c. Spatially uniform photodoping results in chiral domain order[29]. The resistivity of this domain-textured state can drop by 1 to 3 orders of magnitude[23,30,31] with respect to the C state as shown in Fig. 1e. This high-conductance state is metastable over 250 mK < $T$ < 190 K, with a temperature-dependent lifetime[24]. Below 20 K the lifetime is sufficiently long that it may be considered non-volatile. Although the details of the electronic conduction mechanisms are currently still a matter of debate[17,31–33], scanning tunneling spectroscopy (STS) maps show that both the domain walls (DWs) and the domains themselves have gapless electronic spectra[26,27], implying that metallic conductivity involves itinerant electrons in domains *and* in the DWs. Since a myriad of domain configurations are possible, configurational states with different electronic crystal domain structure and different inherent resistivity may be used to store a significant amount of information[34]. Here we opt for the simplest case, using charge injection switching between the high-resistance (HI) domain-free C state and the lowest stable resistance (LO) domain state as the basis of a memory device.

**Microscopic visualization of a Wigner crystal reconfiguration.** To observe the microscopic changes in electronic configurations of polarons corresponding to write (W) and erase (E) switching operations of such a device we use a multiprobe STM at 4.5 K (see Fig. 2a and Methods). Unlike in conventional STM imaging[26] where the current passes vertically through the tip and through the sample to ground, here the current flows laterally between two tips. Two tips (#1 and #2) are used to apply current pulses, and a third tip (#3) is used to image the area where the current flows *between* contacts (Fig. 2a). The V-I curve shown in Fig. 2b describes the experimental protocol. (The shape of the curve is discussed later in the device section, Fig. 4a and in Methods). We start in the HI state and ramp up the W current $I_{12}$ (red triangles) using 50 µs pulses. Above a threshold $I_W^T \sim 4.8$ mA, an abrupt drop of resistance signifies that the sample switches to the LO state. The resistance switching ratio is reduced compared with devices (shown later) because of the $\sim 1 k\Omega$ STM tip contact resistances and because the current is not laterally confined. The corresponding STM images of HI and LO states obtained with tip #3 are shown by the images $\boxed{0}$ and $\boxed{1}$ respectively in Fig. 2c, revealing the characteristic domain pattern of the LO state. To visualize the erase (E) process, we use a sequence of 50 µs current pulses $I_E (= I_{12})$ increasing the magnitude from zero, while monitoring the step-by-step reordering of the charge configuration with tip #3. A selection of the STM images taken after consecutive pulses with increased amplitude is shown in Fig. 2c $\boxed{1} - \boxed{6}$. We see that as $I_E$ increases, the charge configuration is stable, and resistance is nearly ohmic up to a threshold current $I_E^T \simeq 2$ mA (Fig. 2b). Above $I_E^T$ we observe a change of slope, which coincides with an onset of dropping polaron density $\rho$ (Fig. 3a), defined as the number of polarons per nm$^2$. The change of slope near 2 mA thus appears to be associated with the onset of domain wall annihilation. Thereafter DWs gradually disappear (Fig. 2c $\boxed{3} - \boxed{6}$), and above $I_E = 4.4$ mA only one DW remains visible within the field of view. The domain pattern is different each time the experiment is repeated, so pinning by crystal defects, lattice imperfections and impurities do not appear to play a significant role in the switching process, which is important for endurance and reproducibility of practical devices.

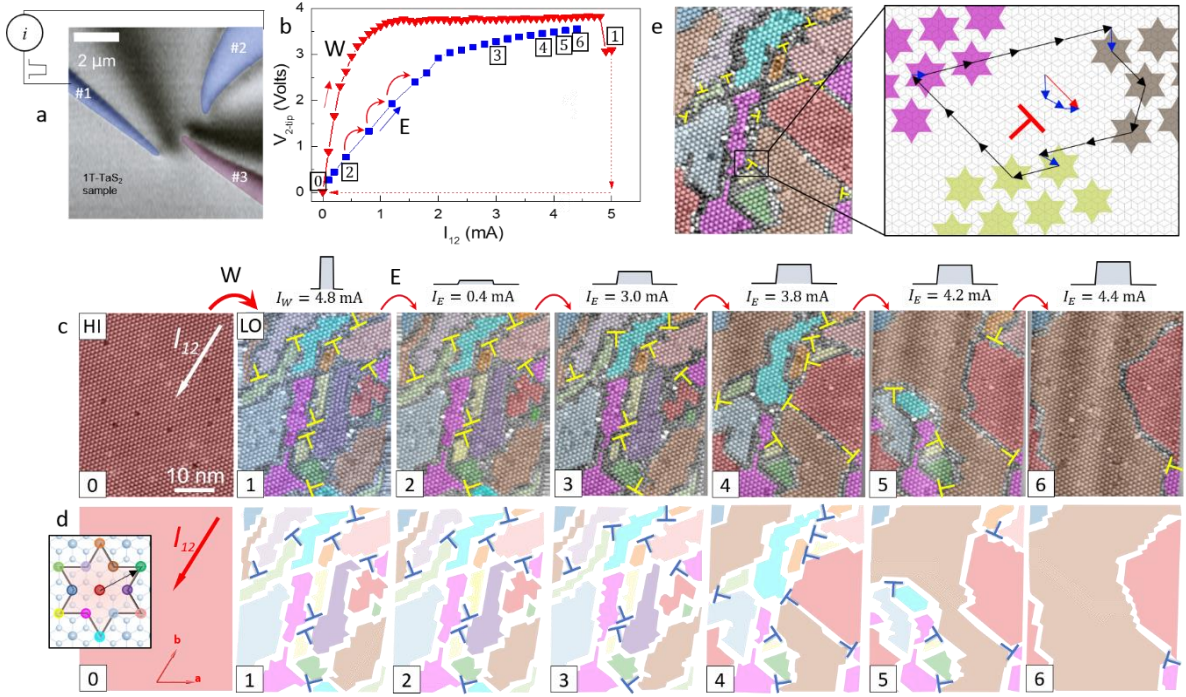

**Figure 2. Visualization of charge domain reconfiguration associated with device operations.** a) An SEM image of the 3-tip setup for sourcing of the current and STM imaging. b) V-I curve during the switching cycle: Initially, switching to the LO state is performed by ramping up the W current from zero up to $I_{12}$=5 mA between tips #1 and #2 (red triangles). Switching from HI to LO state occurs at ~4.8 mA. Erasing back to the HI state is performed by ramping up the E current from zero up (blue squares). c) STM images measured by tip #3; ⓪ and ① show HI and LO state respectively while images ② - ⑥ show the evolution of electronic configuration during the E sequence. The yellow ⊥ symbols represent NTTDs. d) Simplified version of Fig. 2c, showing color-coded domains that signify the displacement of domains relative to the original C structure. The insert to ⓪ shows the color scheme representing 12 possible non-zero displacement vectors which characterize the mutual configurations of polarons in adjacent domains. e) A detailed image of the Burger's vector $\vec{B}$ construction for a NTTD at a crossing of three domain walls (see SI for details).

**Dislocations and metastability of the LO state.** Categorizing the relative displacements of domains relative to each other on the electronic lattice in Fig. 2c by the color code defined by the insert to Fig. 2d, and analyzing the domain-wall junctions, we observe non-trivial topological defects (NTTDs) in the form of dislocations. The importance of these dislocations lies in the fact that they are topologically protected and can be removed only by defect-antidefect pair annihilation, i.e NTTDs with equal and opposite winding numbers. This imparts a metastability to the configurational state of the crystal. Motion of domains and the charges within them is strongly inhibited. The dislocations can be characterized by a Burger's vector construction on the electronic lattice around domain wall crossings[29] (shown by ⊥ dislocation symbols in Figs. 2c, d and e). An example of a domain junction with non-zero *electronic* Burger's vector construction $\vec{B}$ is shown in Fig. 2e (see the SI for a more detailed construction). The evolution of dislocations during an erase sequence is shown by color maps (Fig. 2d). In correspondence to the V-I curve (Fig. 2b), we see that up to ~2 mA, very little motion of the dislocations is detected. Above $I_E = 2.6$ mA, they start to annihilate with each other, and eventually (in the absence of crystal imperfections) disappear upon complete erase $\sum_i \vec{B}_i = 0$. We note that these effects are topologically equivalent to the behaviour of bubble memory on a much larger scale, except that the latter occurs with magnetic rather than electronic domain ordering[35].

By analyzing the sequence of STM images $\boxed{1}$-$\boxed{6}$ for different $I_E$ (Fig. 2c), we can also determine the total density $\rho$ of polarons in the image, the number $N_p$ of polarons moved, and their direction of motion during switching. The change of $\rho$ during the W/E cycles shown in Fig. 3a is consistent with ~10% extra electrons accommodated within domain walls after the W pulse. From Fig. 2d $\boxed{1}$ − $\boxed{3}$, we see that $N$ is small for $I_E < 2$ mA, but increases abruptly above ~3 mA (Figs. 2d $\boxed{3}$-$\boxed{6}$), reaching $N_p \simeq 280$ per frame at 4.4 mA. The dynamics of motion plotted in Figs. 3b and c for different $I_E$ are particularly interesting. The largest number of electron movements at the highest current (4.4 mA) coincides with one of the crystal axis and occurs both parallel *and antiparallel* to the applied current direction $I_{12}$. This observation is significant, because it shows that (i) the switching of charge configurations is not due to a single-particle drift current, but is a collective reconfiguration, (ii) the current is *not* carried by the observed localised charges and (iii) the current cannot be attributed to electronic percolation as in typical memristors. Moreover, we do *not* observe any sliding motion at any $I_E$, indicating that conventional CDW depinning and sliding phenomena that are common in incommensurate CDW systems are not present here. Switching and charge transport mechanisms are thus very different from conventional memristors and thermally-induced phase-change switching devices[30,36–40], and suggest that different electronic bands are involved in the HI and LO state electronic transport. In fact, Hall measurements show that the dominant current-carriers are holes in the HI state and electrons in the LO state[41,42].

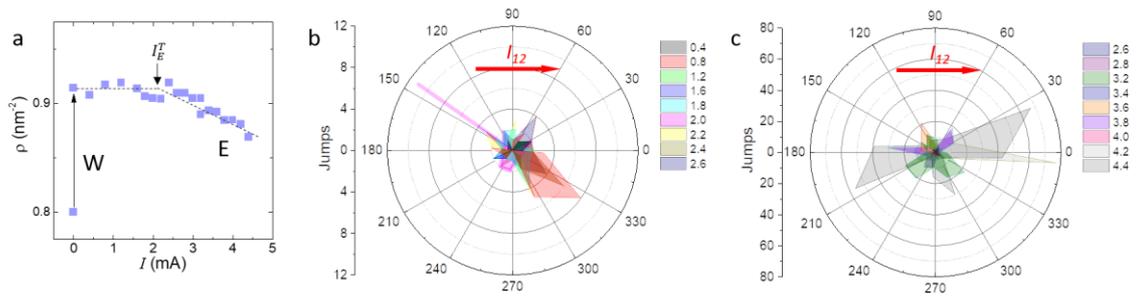

**Figure 3. Polaron dynamics during the E process**. a) The polaron number density $\rho$ (i.e. number of polarons $N_p$ per $nm^2$) within each STM image, counted during the E sequence (see SI for counting method). b) and c) show how many polarons move at each angle for the whole E sequence presented in Fig. 2c. b) For $I_E(= I_{12}) <$ 2.6 mA, the direction of motion appears random. c) For $I_E > 2.6$ mA, the motion *increases*, parallel and antiparallel to the current direction $I_{12}$ indicated by the arrow (note the change of scale from b) to c)).

The average microscopic energy barrier $E_B$ separating the different configurational states can be determined from a measurement of the temperature dependence of the resistance relaxation rate $r = 1/\tau_R$ of the $R_{LO}$ state under small current $I \ll I_E^T$ which follows roughly exponential behavior $R \propto R_{HI}(1 - e^{-t/\tau_R})$. By fitting the time-dependence of thermally-induced resistance relaxation to simple Arrhenius behavior for the relaxation rate $r = r_0 e^{-E_B/kT}$ we obtain $E_B$ (see SI for fits to the T-dependence of $r$). Devices of different sizes and on different substrates give consistent barrier energies in the range $E_B = 15 \sim 21$ meV (180~248 K). These values are smaller than those reported in recent measurements of metastable states induced by electric pulses in 20 $\mu m$ thick crystals at higher temperatures. The differences may indicate that different pinning mechanisms may be relevant for different domain sizes in different temperature ranges, present in crystals of different thickness[43]. A high sensitivity of $E_B$ to strain was also noted previously[24].

The STM experiments show that while the W mechanism is caused by a charge pulse injection into the insulating ($R_{HI}$) state with no domains, the E mechanism involves charge pulse injection into a conducting ($R_{LO}$) state with a dense domain structure. Assuming a constant voltage source applied across the device, the supplied power is significantly larger in the $R_{LO}$ state that the $R_{HI}$ state, whence Joule heating is significantly higher for the E pulse that the W pulse, giving rise to a significant thermal asymmetry in the effect of W and E pulses. Thus, while the W switching is a rapid, non-thermal charge reconfiguration process driven mainly by charge injection[23,24,28,29], the E process shown in Fig. 2 involves a thermally-activated reconfiguration process, which is consistent with the absence of directionality in the dynamics shown in Fig. 3.

E pulses which result in maximum temperatures *below* the CCDW transition temperature $T_{CCDW}$ (190 K), cause thermally-activated domain reconfiguration across the energy barrier $E_B$ (shown in the SI). The relaxation was previously shown to proceed via a discrete sequence of jumps over a "Devil's staircase" of intermediate metastable structures with different degrees of discommensurability[24], eventually reaching the CCDW ($R_{HI}$) state at a rate given by *r*. Complete, rapid erase requires that the sample is heated momentarily *above* $T_{CCDW}$. Empirically, it has been shown that the $R_{LO}$ state can be erased with various protocols and pulse sequences. Here we have used single E pulse lengths of 50 μs (Fig. 2), 20 ms (Fig. 4f) and 600 ms (Fig. 4a, b). Previously, an E pulse train of 50 ps optical pulses at 5 μs intervals was reported[23], suggesting that fast E processes are possible, but to determine to what extent the process is thermal requires further investigations.

The demonstrated electronic domain configuration mechanism in Fig. 2 has important consequences for practical device characteristics in terms of energy, speed and data retention. To investigate the W switching energy and speed we replace the STM tip electrodes with nanofabricated ones. This way the geometry of the device is better defined, and transmission line contacts can be made for short pulse measurements. The V-I curve for a W cycle at 20 K is shown in Fig. 4a for *incrementally* increasing current pulses $I_W$ through the device with an inter-contact distance $L = 145$ nm shown in the insert.

Here the pulses are intentionally long of length $\tau_W = 600$ ms, to allow long instrumental time constants and hence low noise in the measurement. We observe a non-linear V-I curve with an initial slope $R_{HI}$ (dashed $R_{HI}$ line) up to the $I_W^T \simeq 0.165$ mA. Above this current the voltage drops from 0.65 V to 0.08 V and the device switches to a linear V-I relation with resistance $R_{LO}$ (dashed $R_{LO}$ line). The V-I curve for W can be fit with $V = V_0 \ln(I/I_0 + 1)$ for $I < I_W^T$ (green line), and $V = R_{LO} I$ for $I > I_W^T$ (dashed $R_{LO}$ line) in agreement with ref.[25]. This behavior is consistent with the STM tip switching shown in Fig. 2b, taking into account different current paths in the devices. An E pulse with a higher threshold $I_E^T > I_W^T$ is used to switch back to the pristine state. The cycle of erase pulses is illustrated in Fig. 4b where the current is ramped from $I = 0$ to 0.4 mA. Above an erase-threshold $I_E^T \simeq 0.25$ mA the system is seen to switch to a high resistance state, and the device reverts back to $R_{HI}$ as current is lowered to zero. In the next two sections we investigate the scaling of the device operation parameters (pulse length, switching voltage and switching energy) that are important for practical applications.

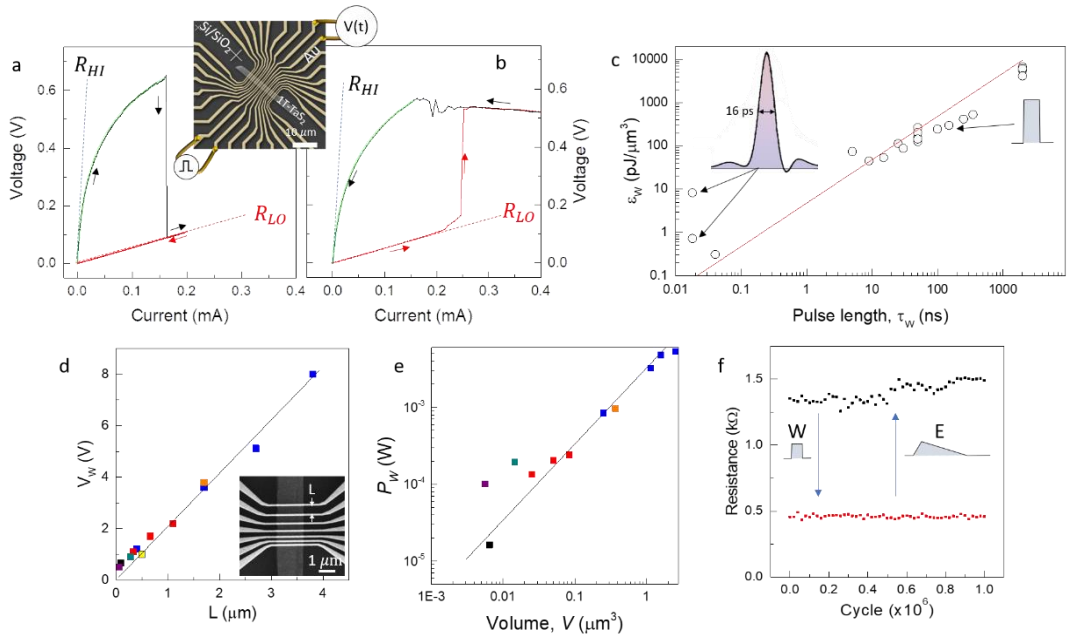

**Figure 4. Fabricated device characteristics at 20 K.** a) and b) show the V-I curves for the W and E processes respectively. The green line is a fit to the data in both cases (see text). The insert to a) shows a test device with 8 cells. c) The switching energy density $\varepsilon_W$ as a function of pulse length $\tau_W$. The insert shows the actual pulse shapes. d) The switching threshold voltage $V_W$ as a function of distance between the electrodes L measured on devices like the one shown in the insert. e) The switching power $P_w$ as a function of device volume $\mathcal{V}$. Different symbol colors are for different physical devices. f) An endurance measurement showing cycling between $R_{LO}$ and $R_{HI}$ for $10^6$ cycles.

**Switching speed.** On the basis of the non-thermal configurational W mechanism we may expect switching to be very fast. Using appropriate transmission line contacts on Si/SiO$_2$ substrates (see insert to Fig. 4a, Methods), we systematically change the W pulse length $\tau_W$ and measure the switching energy density $\varepsilon_W = E_W/\mathcal{V}$ (here $\mathcal{V}$ is the crystal volume between electrodes, and $E_W$ is the switching energy of the device). The results are shown in Fig. 4c over > 5 orders of magnitude 16 ps $< \tau_W <$ 2 μs. The line shows a fit to a linear relation $\varepsilon(\tau_W) = \gamma \tau_W$, with $\gamma \simeq 3.3 \frac{\text{pJ}}{\text{ns μm}^3}$.

The lowest value of W pulse energy $E_W$ is obtained with a $\tau_W = 16$ ps FWHM pulse, (the insert to Fig. 4c shows the actual 11 ps rise-time pulse measurement). A device with L=150 nm, $R_{HI} = 8200$ Ω, $R_{LO} = 1800$ Ω and $V_W = 1\,V$, and is measured at 20 K to give $E_W = 2.2$ fJ, where the resistance was measured before and after switching, and $\tau_W$ was used to calculate $E_W$. The value of 2.2 fJ is extremely small compared with other current memory devices, but is still significantly larger than the measured *microscopic* barrier $E_B = 15 \sim 21$ meV ($2.4 \sim 3.4 \times 10^{-6}$ fJ), which implies that devices with much smaller $E_W$ can be built before reaching fundamental limits. It is instructive to compare $E_W$ and $E_B$ with the lowest possible energy difference between two states that can be discerned thermodynamically on the basis of their entropy difference $\Delta E_L = k_B T ln2$ (the Brillouin-Landauer thermodynamic limit[44]). At $T = 4$ K, $\Delta E_L \simeq 3.8 \times 10^{-8}\,fJ$, so the current CCM devices are still far from this limit. The observed scaling laws suggest that $E_W$ can be reduced further by reducing $L$ and $\tau_W$. Deviations from linearity might be expected when $\tau_w$ approaches the intrinsic switching time $\tau_s = 0.45$ ps, or when a competing mechanism starts to consume significant energy, such as heating. These effects are likely to be important for device optimisation, particulary for erase protocols.

**Scaling of switching voltage and power with device size.** In accordance with expectations based on the STM measurements, the switching threshold voltage $V_W$ should scale with distance between contacts $L$. The result of a measurement is shown in Fig. 4d, with linear scaling of $V_W$ over nearly two orders of magnitude 60 nm $< L <$ 4 μm. Similarly, the switching power defined as $P_W = I_W V_W$ as a function of device volume $\mathcal{V}$ is also linear over 3 orders of magnitude extending to $6 \times 10^{-3}$ μm$^3$ (Fig. 4e). A fit gives the switching power density at threshold: $p_W = \frac{P_W}{\mathcal{V}}, \simeq 3 \times 10^{-3}$ W/μm$^3$. With 16 ps switching pulses and a practically realizable $10 \times 10$ nm device on a 30 nm thick TaS$_2$ film the switching energy is $E_W = \gamma \tau_W \mathcal{V} \sim 10^{-4}$ fJ, which corresponds to a switching power $P_W = p_W \mathcal{V} \sim 5$ nW.

For practial applications it is interesting to investigate switching reproducibility and stability of the resistance in the switched state. In Fig. 4f we show a typical endurance measurement of a relatively large device with $L = 300$ nm and very long pulses ($V_W = 0.45$ V, $\tau_w = 2.7$ μs square wave W pulses, $V_E = 0.43$ V, $\tau_E = 20$ ms asymmetric triangular E pulses). Each pair of corresponding $R_{HI}$ and $R_{LO}$ values represents a measurement done after 20k cycles of W and E pulses. Surprisingly, $R_{LO}$ is remarkably stable throughout the whole $10^6$ cycles, while $R_{HI}$ increases slightly and later stabilizes. A measurement of $R_{LO}$ for 4800 R cycles shows the LO state is stable as long as the device is kept below 20 K (See SI for plot). This is consistent with the extrapolated lifetime of the $R_{LO}$ state at 4K, based on the measured activation energy presented in the supplement, of $\sim 10^{14}$ years.

**Discussion**

Briefly comparing the measured operating parameters of CCM with leading magnetic random access memory (MRAMs) and PCM devices, the smallest energy/bit values so far reported by voltage-controlled magnetic anisotropy switching in multilayer IrMn/CoFeB/MgO and Ta/CoFeB/MgO magnetic tunnel junction (MTJ) devices are at 39 ~ 40 fJ/bit[45]. The lowest theoretically predicted value of a few fJ/bit[2], comparable to the 2.2 fJ presently reported, and much higher than $10^{-4}$ fJ predicted for CCM in the previous paragraph. Phase change memories have both significantly higher switching energies (> 2.5 pJ) and switching voltages (> 1 V[46,47]) than CCM, and even the smallest memristors such as 6 nm crossbar memristor arrays have a write power of $P_W = 0.23$ μW[48], which is significantly higher than ~3 nW for CCM. We note that our demonstration of a man-made *non-volatile* electronic memory device uses less energy per bit (2.2 fJ/bit) than a human brain (10 fJ/synapse), and potentially much less than artificial synapses[49]. The CCM's advantage is that it's > 9 orders of magnitude faster. (A comparison of different devices is summarized in the SI.)

With demonstrated scalability in size (60 nm − 4 μm), pulse length (16 ps − 600 ms) and voltage (0.3 V − 10 V) and with a wide operating temperature (250 mK − 200 K) the CCM devices appear to be very versatile. The operating temperature range at high T could be extended by appropriate choice of substrate[24], but so far no other electronically ordered material has been found to exhibit significant metastability. While we have demonstrated that the basic mechanism of switching is based on charge reconfiguration, the role of heating that inevitably accompanies charge injection still needs to be understood. Surprisingly, the device variability is small (as seen from the data on multiple devices shown in the scaling plots). Importantly for applications, thin films of 1T-TaS$_2$ can be grown by various means[50–52] promising technological flexibility. An immediate application of the device could be in cryocomputing, which has been heralded as an obvious solution to the overall challenge of reducing dissipation[9,53]. In spite of huge research efforts and availability of superconducting circuits performing both single flux quantum (SFQ) and quantum information processing[9], the absence of a fast low-energy cryogenic memory has prevented significant upscaling[53], so CCM may offer a possible breakthrough. For interfacing CCM with single flux quantum devices, which is an obvious target application, nanocryotrons have been demonstrated to be an excellent match in terms of output voltage, speed and impedance[54,55]. In comparison to complementary metal-oxide semiconductor (CMOS) devices[56,57] and superconducting loop memories[58], CCM offers significant advantages in terms of scaling, size, speed, energy/bit and operational simplicity. Irrespective of application and path to fabrication, the ultrafast CCM devices introduce a new concept that challenges existing information storage paradigms.

## *Methods*

The 1T-TaS$_2$ single crystals were synthesized using the vapor phase transport method. More than 1000 devices (such as in the insert to Fig.4a) were fabricated by deposition of typically 30 − 100 nm thick single crystals on Si/SiO$_2$ substrates with 80 nm thick sputtered Au contacts over a 5 nm Au/Pd contact layer using e-beam lithography. The contact resistances with PdAu/Au electrodes, obtained from a 4-probe measurement were typically $100 - 200$ Ω. DC transport device measurements were performed using a standard four-point contact technique. For $\tau_W > 4$ ns an electrical signal generator (Siglent SDG 1050) was used, while for $\tau_W < 100$ ps electrical pulses were generated using an amplified SYMPULS signal generator giving $\pm 2.2$ V at $\tau = 16$ ps (FWHM, 11 ps rise-time at the input to the cryostat after amplifier and bias tee). For frequencies > 1 GHz, transmission lines were used throughout, such as shown in the insert to Fig. 4a and 4d. The samples for all STM measurements were cleaved in UHV. The resistance measurements were performed after switching by measuring and fitting I-V curves by 2 or 4-contact method, as indicated. The 2-tip and 4-tip STM was performed in an Omicron 4-probe LT at 4 K. For V-I measurements, with two outer tungsten tips $1 - 3$ μm apart we first made tunneling contact, and then advanced the tips into the sample until an approximately linear I-V curve was reached, (typically ~40 nm beyond the contact point). For 4-contact measurements, two inner tips are used to measure the voltage drop $V_{23}$ at two points in the area between the current sourcing tips. By comparing 2-tip and 4-tip measurements, we determine the contact resistance to be ~1 kΩ, while the intrinsic sample resistance between the inner tips ~ 1 μm apart is $R_{12} \sim 40$ Ω. Compared with a fabricated device, the current path between STM tips is not laterally confined, and heating is higher because of the higher tip-sample contact resistance ~1 kΩ, vs. ~100 Ω for typical fabricated contacts.

**Author contributions.** AM, RV, IAM, VS and IV performed electrical measurements experiments and AM, D. Svetin, DK, JR and YE fabricated devices, YG, MD, JR, AK and AM performed STM experiments; MD'A, D. Stornaiulo and FT performed the 250 mK switching experiment; AM, D. Svetin, YG and DM devised the experiments, and DM wrote the paper.

**Acknowledgments**. This project has received funding from the the EU-H2020 research and innovation programme under grant agreement No 654360 having benefited from the access provided by Paul Scherrer Institute in Villigen, Switzerland within the framework of the NFFA-Europe Transnational Access Activity. We wish to acknowledge the help of L. Cindro from F9 at JSI on contact bonding. We thank for the support from the Slovenian Research Agency (P1-0040, A.M. to PR-08972, A.K. to PR-06158, J.R. to PR-07589, D.S. to I0-0005), Slovene Ministry of Science (Raziskovalci-2.1-IJS-952005), ERC AdG (GA320602) and ERC PoC (GA7677176). We thank the CENN Nanocenter for the use of an AFM and the LDI. This project has received funding from the European Union's Horizon 2020 research and innovation program under the Marie Skłodowska-Curie grant agreement No 701647.

# *References*

1. Parkin, S. S. P., Hayashi, M. & Thomas, L. Magnetic Domain-Wall Racetrack Memory. *Science* **320**, 190–194 (2008).
2. Wang, L. *et al.* Voltage-Controlled Magnetic Tunnel Junctions for Processing-In-Memory Implementation. *IEEE Electron Device Lett.* **39**, 440–443 (2018).
3. Peng, S. *et al.* Field-Free Switching of Perpendicular Magnetic Tunnel Junction via Voltage-Gated Spin Hall Effect for Low-Power Spintronic Memory. arXiv:1804.11025 [physics.app-ph] (2018)
4. Amiri, P. K. *et al.* Electric-Field-Controlled Magnetoelectric RAM: Progress, Challenges, and Scaling. *IEEE Trans. Magn.* **51**, 1–7 (2015).
5. Junquera, J. & Ghosez, P. Critical thickness for ferroelectricity in perovskite ultrathin films. *Nature* **422**, 506–509 (2003).
6. Sun, W. *et al.* Understanding memristive switching via in situ characterization and device modeling. *Nat. Commun.* **10**, 3453 (2019).
7. Wang, M. *et al.* Field-free switching of a perpendicular magnetic tunnel junction through the interplay of spin–orbit and spin-transfer torques. *Nat. Electron.* **1**, 582–588 (2018).
8. Guo, L. A Silicon Single-Electron Transistor Memory Operating at Room Temperature. *Science* **275**, 649–651 (1997).
9. Anders, S. *et al.* European roadmap on superconductive electronics – status and perspectives. *Phys. C Supercond.* **470**, 2079–2126 (2010).
10. Janod, E. *et al.* Resistive Switching in Mott Insulators and Correlated Systems. *Adv. Funct. Mater.* **25**, 6287–6305 (2015).
11. Wilson, J. A., Di Salvo, F. J. & Mahajan, S. Charge-density waves and superlattices in the metallic layered transition metal dichalcogenides. *Adv. Phys.* **24**, 117–201 (1975).
12. Vodeb, J. *et al.* Configurational electronic states in layered transition metal dichalcogenides. *New J. Phys.* **21**, 083001 (2019).
13. Sipos, B. *et al.* From Mott state to superconductivity in 1T-$TaS_2$. *Nat. Mater.* **7**, 960–965 (2008).
14. Rohwer, T. *et al.* Collapse of long-range charge order tracked by time-resolved photoemission at high momenta. *Nature* **471**, 490–493 (2011).
15. Cho, D., Cho, Y.-H., Cheong, S.-W., Kim, K.-S. & Yeom, H. W. Interplay of electron-electron and electron-phonon interactions in the low-temperature phase of 1T−$TaS_2$. *Phys. Rev. B* **92**, 085132 (2015).
16. Spijkerman, A., de Boer, J. L., Meetsma, A., Wiegers, G. A. & van Smaalen, S. X-ray crystal-structure refinement of the nearly commensurate phase of 1T−$TaS_2$ in (3+2) -dimensional superspace. *Phys. Rev. B* **56**, 13757–13767 (1997).
17. Ritschel, T. *et al.* Orbital textures and charge density waves in transition metal dichalcogenides. *Nat. Phys.* **11**, 328–331 (2015).
18. Dai, H., Chen, H. & Lieber, C. M. Weak pinning and hexatic order in a doped two-dimensional charge-density-wave system. *Phys. Rev. Lett.* **66**, 3183–3186 (1991).
19. Karpov, P. & Brazovskii, S. Modeling of networks and globules of charged domain walls observed in pump and pulse induced states. *Sci. Rep.* **8**, 4043 (2018).
20. Ritschel, T., Berger, H. & Geck, J. Stacking-driven gap formation in layered 1T-$TaS_2$. *Phys. Rev. B* **98**, 195134 (2018).
21. Qiao, S. *et al.* Mottness Collapse in 1T−$TaS_{2–x}Se_x$ Transition-Metal Dichalcogenide: An Interplay between Localized and Itinerant Orbitals. *Phys. Rev. X* **7**, 041054 (2017).
22. McMillan, W. L. Theory of discommensurations and the commensurate-incommensurate charge-density-wave phase transition. *Phys. Rev. B* **14**, 1496–1502 (1976).
23. Stojchevska, L. *et al.* Ultrafast Switching to a Stable Hidden Quantum State in an Electronic Crystal. *Science* **344**, 177–180 (2014).
24. Vaskivskyi, I. *et al.* Controlling the metal-to-insulator relaxation of the metastable hidden quantum state in 1T-$TaS_2$. *Sci. Adv.* **1**, e1500168 (2015).
25. Vaskivskyi, I. *et al.* Fast electronic resistance switching involving hidden charge density wave states. *Nat. Commun.* **7**, 11442 (2016).
26. Ma, L. *et al.* A metallic mosaic phase and the origin of Mott-insulating state in 1T-$TaS_2$. *Nat. Commun.* **7**, 10956 (2016).

27. Cho, D. *et al.* Nanoscale manipulation of the Mott insulating state coupled to charge order in 1T-TaS$_2$. *Nat. Commun.* **7**, 10453 (2016).
28. Ravnik, J., Vaskivskyi, I., Mertelj, T. & Mihailovic, D. Real-time observation of the coherent transition to a metastable emergent state in 1T−TaS$_2$. *Phys. Rev. B* **97**, 075304 (2018).
29. Gerasimenko, Y. A., Karpov, P., Vaskivskyi, I., Brazovskii, S. & Mihailovic, D. Intertwined chiral charge orders and topological stabilization of the light-induced state of a prototypical transition metal dichalcogenide. *Npj Quantum Mater.* **4**, 32 (2019).
30. Yoshida, M. *et al.* Controlling charge-density-wave states in nano-thick crystals of 1T-TaS$_2$. *Sci. Rep.* **4**, 7302 (2014).
31. Svetin, D., Vaskivskyi, I., Brazovskii, S. & Mihailovic, D. Three-dimensional resistivity and switching between correlated electronic states in 1T-TaS$_2$. *Sci. Rep.* **7**, 46048 (2017).
32. Butler, C. J., Yoshida, M., Hanaguri, T. & Iwasa, Y. Mottness versus unit-cell doubling as the driver of the insulating state in 1T-TaS$_2$. *Nat. Commun.* **11**, 2477 (2020).
33. Park, J. W., Cho, G. Y., Lee, J. & Yeom, H. W. Emergent honeycomb network of topological excitations in correlated charge density wave. *Nat. Commun.* **10**, 4038 (2019).
34. Ma, Y., Wu, D., Lu, C. & Petrovic, C. The electric pulses induced multi-resistance states in the hysteresis temperature range of 1T-TaS$_2$ and 1T-TaS$_{1.6}$Se$_{0.4}$. *Appl. Phys. Lett.* **116**, 171906 (2020).
35. Seul, M. & Murray, C. A. Scale transformation of magnetic 'bubble' arrays: coupling of topological disorder and polydispersity. *Science* **262,** 558 (1993).
36. Yu, Y. *et al.* Gate-tunable phase transitions in thin flakes of 1T-TaS$_2$. *Nat. Nanotechnol.* **10**, 270–276 (2015).
37. Yoshida, M., Suzuki, R., Zhang, Y., Nakano, M. & Iwasa, Y. Memristive phase switching in two-dimensional 1T-TaS$_2$ crystals. *Sci. Adv.* **1**, e1500606 (2015).
38. Tsen, A. W. *et al.* Structure and control of charge density waves in two-dimensional 1T-TaS$_2$. *Proc. Natl. Acad. Sci.* **112**, 15054–15059 (2015).
39. Hollander, M. J. *et al.* Electrically Driven Reversible Insulator–Metal Phase Transition in 1T-TaS$_2$. *Nano Lett.* **15**, 1861–1866 (2015).
40. Geremew, A. K. *et al.* Bias-Voltage Driven Switching of the Charge-Density-Wave and Normal Metallic Phases in 1T-TaS$_2$ Thin-Film Devices. *ACS Nano* **13**, 7231–7240 (2019).
41. Yoshida, M., Gokuden, T., Suzuki, R., Nakano, M. & Iwasa, Y. Current switching of electronic structures in two-dimensional 1T−TaS$_2$ crystals. *Phys. Rev. B* **95**, 121405 (2017).
42. Inada, R., Ōnuki, Y. & Tanuma, S. Hall effect of 1T-TaS$_2$ and 1T-TaSe$_2$. *Phys. B+C* **99**, 188–192 (1980).
43. Ma, Y. *et al.* Observation of multiple metastable states induced by electric pulses in the hysteresis temperature range of 1T−TaS$_2$. *Phys. Rev. B* **99**, 045102 (2019).
44. Landauer, R. Irreversibility and Heat Generation in the Computing Process. *IBM journal 5*, 183-191 (1961).
45. Dorrance, R. *et al.* Diode-MTJ Crossbar Memory Cell Using Voltage-Induced Unipolar Switching for High-Density MRAM. *IEEE Electron Device Lett.* **34**, 753–755 (2013).
46. Loke, D. *et al.* Breaking the Speed Limits of Phase-Change Memory. *Science* **336**, 1566–1569 (2012).
47. Ding, K. *et al.* Phase-change heterostructure enables ultralow noise and drift for memory operation. *Science* **366**, 210–215 (2019).
48. Pi, S. *et al.* Memristor crossbar arrays with 6-nm half-pitch and 2-nm critical dimension. *Nat. Nanotechnol.* **14**, 35–39 (2019).
49. Xu, W., Min, S.-Y., Hwang, H. & Lee, T.-W. Organic core-sheath nanowire artificial synapses with femtojoule energy consumption. *Sci. Adv.* **2**, e1501326 (2016).
50. Enomoto, H., Kawano, T., Kawaguchi, M., Takano, Y. & Sekizawa, K. Van der Waals Growth of Thin TaS$_2$ on Layered Substrates by Chemical Vapor Transport Technique. *Jpn. J. Appl. Phys.* **43**, L123–L126 (2004).
51. Sanders, C. E. *et al.* Crystalline and electronic structure of single-layer TaS$_2$. *Phys. Rev. B* **94**, 081404 (2016).
52. Wang, X. *et al.* Chemical Growth of 1T-TaS$_2$ Monolayer and Thin Films: Robust Charge Density Wave Transitions and High Bolometric Responsivity. *Adv. Mater.* **30**, 1800074 (2018).

53. Holmes, D. S., Ripple, A. L. & Manheimer, M. A. Energy-Efficient Superconducting Computing—Power Budgets and Requirements. *IEEE Trans. Appl. Supercond.* **23**, 1701610–1701610 (2013).
54. Zhao, Q.-Y., McCaughan, A. N., Dane, A. E., Berggren, K. K. & Ortlepp, T. A nanocryotron comparator can connect single-flux-quantum circuits to conventional electronics. *Supercond. Sci. Technol.* **30**, 044002 (2017).
55. Mraz, A., Kabanov, V. V. & Mihailovic, D. Nanocryotron-driven Charge Configuration Memory Device. (To be published)
56. Van Duzer, T. *et al.* 64-kb Hybrid Josephson-CMOS 4 Kelvin RAM With 400 ps Access Time and 12 mW Read Power. *IEEE Trans. Appl. Supercond.* **23**, 1700504–1700504 (2013).
57. Mukhanov, O. A., Kirichenko, A. F., Filippov, T. V. & Sarwana, S. Hybrid Semiconductor-Superconductor Fast-Readout Memory for Digital RF Receivers. *IEEE Trans. Appl. Supercond.* **21**, 797–800 (2011).
58. Murphy, A., Averin, D. V. & Bezryadin, A. Nanoscale superconducting memory based on the kinetic inductance of asymmetric nanowire loops. *New J. Phys.* **19**, 063015 (2017).

# *Energy efficient manipulation of topologically protected states in non-volatile ultrafast charge configuration memory devices*

## *Supplementary information*

### *1. Trivial and non-trivial domain wall defect constructions on a sparsely filled hexagonal lattice*

A Burgers' vector construction is made around domain wall intersections on the primitive crystal lattice (grey), which is derived from an STM image (Supplementary Figure 1). The black arrows represent electron lattice vectors, while the blue arrows represent crystal lattice vectors.

In the left example a trivial defect has Burger's vector sum $\boldsymbol{B} = 0$. The right example is non-trivial ($\boldsymbol{B} \neq 0$), and has $\boldsymbol{B} = \boldsymbol{A}_C = 3\boldsymbol{a} + \boldsymbol{b}$, where $\boldsymbol{a}$ and $\boldsymbol{b}$ are the atomic lattice vectors, and $\boldsymbol{A}_C$ is the primitive electronic lattice vector of the commensurate (HI) state. The latter is topologically protected and can annihilate only with a defect with an equal and opposite $\boldsymbol{B}$.

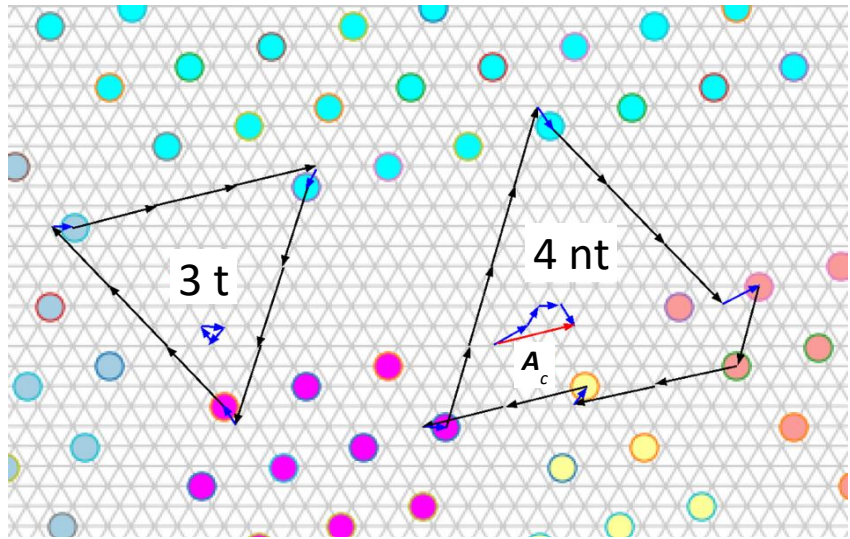

**Supplementary Figure 1:** Burger's vector constructions for 3-domain trivial (3t) and 4-domain non-trivial (4nt) domain wall crossings. The colored dots represent the polarons in the STM image corresponding to different domains. The blue arrows are atomic lattice vectors, while the black arrows are the electronic lattice vectors. The Burger's sum of atomic lattice vectors around a vertex are shown with blue arrows. In the case of the 4 nt, the sum is equal to one electornic lattice vector $\boldsymbol{A}_c$.

## 2. Polaron counting

This is done by comparing pairs of images of the same scanned area before and after applying the current impulse. The comparison is made in Python environment and consists in counting the number of polarons which have moved between the two images and the directions of these movements.

To achieve the result, we first need to make sure that the pair of images we want to compare shows the exact same sample's area. While scanning with the STM, indeed, the sample can drift of a few nm and the relative tip-sample position can change. This results in slightly different scanned areas for the two images. In order to cut out this effect, we choose a specific fixed point which remains the same for both the two images (e.g. a specific stationary polaron). Then we translate the two images so that the chosen fixed point results in the same spatial coordinates for both the images. Finally we cut the images' borders in such a way that the two images show exactly the same scanned area.

The next step is the detection of the polarons' centres. For every image our program detects the centre of each polaron exploiting a *skimage* function based on the blobs' recognition by using the Laplacian of Gaussian method. Then we collect all the coordinates of these centres in a pattern of polarons' positions. Supplementary Figure 2a shows two examples of these patterns placed above the corresponding sample's area. Coloured dots represent the polarons' centres.

By comparing the patterns, we can count the number of polarons' movements. In order to do so, let's focus on the David's star which forms the periodic lattice, shown in Supplementary Figure 2b. We can label the position *r* of a single Ta atom on the surface as:

$$r = n\,A + m\,B + a_i + b_j$$

where *A* and *B* are the vectors which connect the different Ta atoms in the centre of the polarons, *n* and *m* are integer numbers which select a specific polaron's centre, while $a_i$ and $b_j$ (i, j = 1, …, 6) are the vectors which connect the centre of the David's star to the Ta atoms on its vertexes and its corners, as shown in figure 2. Quantitatively, $|a_i| = \sqrt{3}\,|b_j| = \sqrt{3} \cdot 0.33$ nm.

Each polaron's centre is uniquely labeled by a pair (*n*, *m*). Any translation which moves the polaron to a new pair (*n'*, *m'*) would make no difference, bringing the system to an identical situation, thanks to the periodicity of the lattice. The only possible meaningful physical translations are those equal to $|a_i|$ or $|b_j|$.

Let's move back to the detection of movements. If we compare the pair of polarons' patterns, we can calculate the distances between each point of the first pattern and its nearest point in the second pattern. These two points are the same polaron's centre in the two different images. Let's call ($x_1$, $y_1$) the coordinates of a specific polaron in the first pattern and ($x_2$, $y_2$) the coordinates of the same polaron in the second image. Thus, the polaron's movement is given by the distance *d*:

$$d = \sqrt{[(x1\text{-}x2)^2 + (y1\text{-}y2)^2]}$$

As we said, the minimum meaningful translations is the one equal to $|b_j|$ = 0.33 nm. This means that the polaron is considered moved only if *d* is larger than 0.33 nm (with a chosen uncertainty of 0.03 nm).

So, by repeating this calculation for every polaron of the pattern, we can count the number of polarons which have moved between the two images. To see this, we can look at Supplementary Figure 2c, where the two patterns of Supplementary Figure 2a are superimposed. When a polaron didn't move, its centre stays in the same coordinates for both the patterns and thus its dot belonging to the first pattern is covered by the other one. When instead the polaron has moved, the two centres don't coincide and both the dots are visible.

Finally we want the direction of the polarons' movements. This is achieve by calculating the shift vectors related to all polarons' movements. If a polaron has moved of a distance $dx = x2 - x1$ along the image's horizontal axis and of a distance $dy = y2 - y1$ along the image's vertical axis, then the angle in respect to the horizontal axis is simply given by *arctan*(*dy*/*dx*). After calculating all the angles of the polarons' movements, we plot them in a polar histogram with 13 bins.

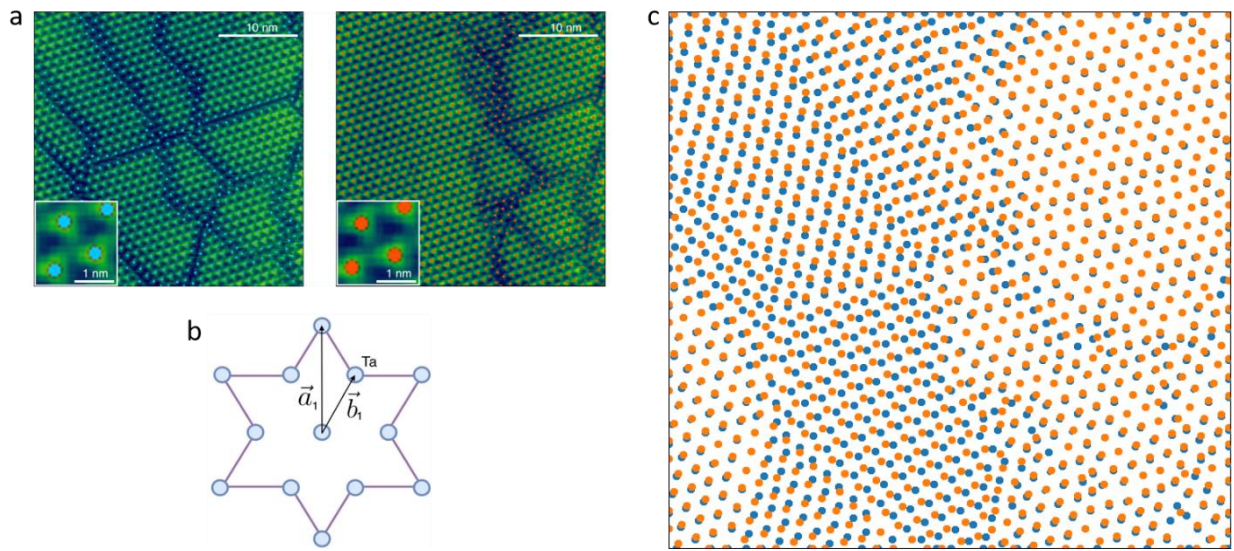

**Supplementary Figure 2: Counting of polarons.** a) Images of two different configurational states with the centers of polarons marked. b) David's star which forms the periodic lattice. c) Superposition of two different configurational states showing the displacements of polaron's centers.

## 3. Microscopic barrier measurement

The microscopic energy barrier $E_B$ separating the different configurational states can be determined from a measurement of the temperature dependence of the resistance relaxation time $\tau_R$ of the $R_{LO}$ state. Irrespective of mechanism, an Arrhenius plot of the relaxation rate with temperature gives the effective barrier energy $E_B$. Two data sets of resistance vs. time at different temperatures for two very different size devices with $l = 4$ µm and $l = 60$ nm fabricated on $SiO_2$ (quartz) and $Si/SiO_2$ (standard oxidized Si) substrates respectively are shown in Figs. 3 a and b. Fits to $R(t)$, using simple exponentials $R(t) = R_C\left(1 - exp\left(-\frac{t}{\tau_R}\right)\right)$, gives approximate Arrhenius-like behavior for $1/\tau_R$ with temperature as shown in Fig. 3c with barrier energies $E_B = 21 \pm 2.3\ meV\ (248 \pm 27\ K)$ and $15.4 \pm 3.4\ meV\ (179 \pm 40K)$ for the $4\ \mu m$ and 60 nm devices respectively. Fig. 3d shows the stability of the written LO state over 4800 Read cycles spread over a two hour long measurement at 20 K.

A more detailed treatment of configurational state relaxation involves a sequence of intermediate metastable states via discrete jumps over a "Devil's staircase", which is better described in terms of stretch exponentials (ref. 24 of main text). The justification for using a simple thermally activated process here is the fact that it is useful for estimating the microscopic barrier energy.

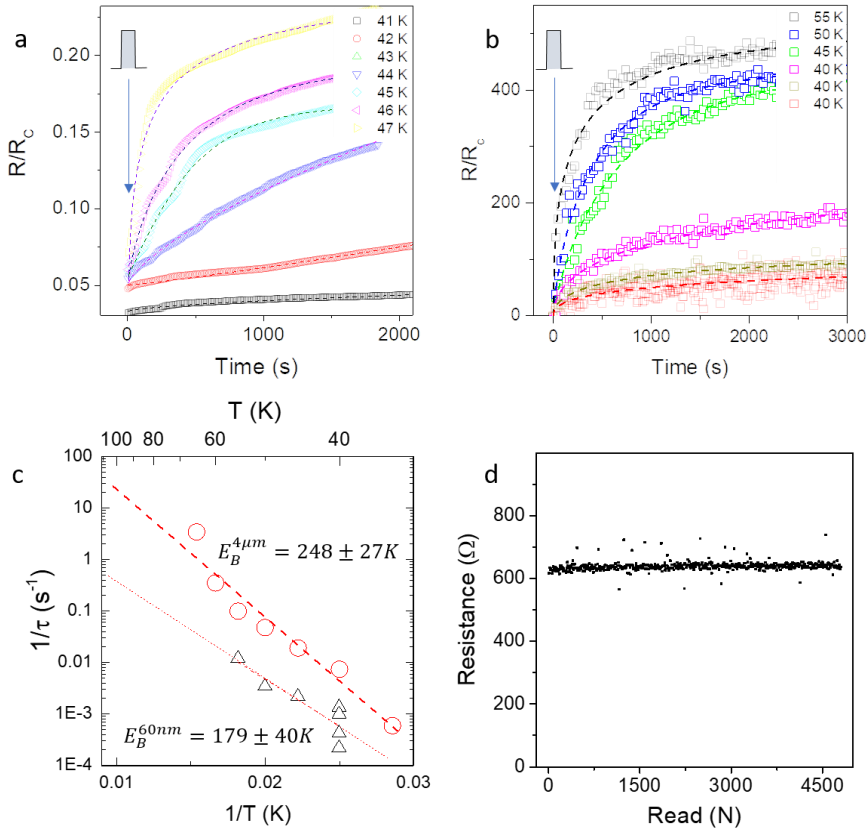

**Supplementary Figure 3: The microscopic energy barrier.** a) and b) The resistivity as a function of time at different temperatures for a 4 µm and 60 nm device respectively. c) The relaxation rate as a function of inverse temperature. The dashed lines are fits to the data, giving $\mathbf{E_B = 248 \pm 27K}$ and $\mathbf{E_B = 179 \pm 40K}$ respectively. d) 4800 Read cycles of written LO state at 20 K.

## 4. *A comparison of leading memory devices – with references*

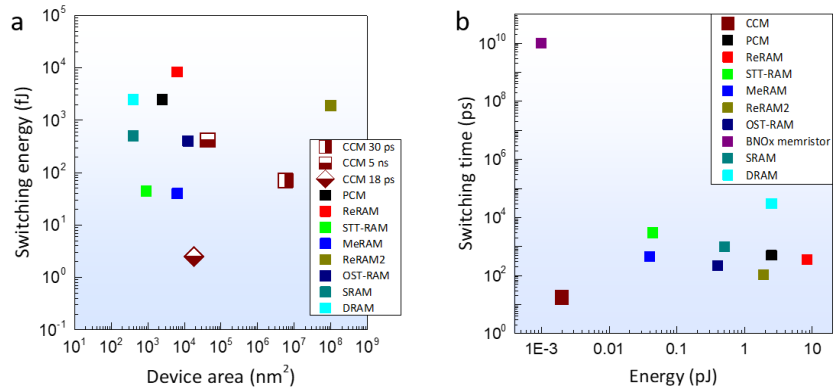

**Supplementary Figure 4: Measured switching energy $E_W$ and speed of leading memory devices.** a) Switching energy in correlation with device area, b) Switching times $\tau_W$ plotted against switching energy. References: PCM[1], ReRAM₁[2], STT-RAM[3], MeRAM[4], ReRAM₂[5], OST-RAM[6], BNO$_x$[7], SRAM[8,9], DRAM[9]

## *References*

1. Loke, D. *et al.* Breaking the Speed Limits of Phase-Change Memory. *Science* **336**, 1566–1569 (2012).
2. Havel, V. *et al.* Ultrafast switching in Ta2O5-based resistive memories. in *2016 IEEE Silicon Nanoelectronics Workshop (SNW)* 82–83 (IEEE, 2016). doi:10.1109/SNW.2016.7577995.
3. Iwata-Harms, J. M. *et al.* Ultrathin perpendicular magnetic anisotropy CoFeB free layers for highly efficient, high speed writing in spin-transfer-torque magnetic random access memory. *Sci. Rep.* **9**, 19407 (2019).
4. Khalili Amiri, P. *et al.* Electric-Field-Controlled Magnetoelectric RAM: Progress, Challenges, and Scaling. *IEEE Trans. Magn.* **51**, 1–7 (2015).
5. Torrezan, A. C., Strachan, J. P., Medeiros-Ribeiro, G. & Williams, R. S. Sub-nanosecond switching of a tantalum oxide memristor. *Nanotechnology* **22**, 485203 (2011).
6. Krivorotov, I. N. *et al.* Ultrafast spin torque memory based on magnetic tunnel junctions with combined in-plane and perpendicular polarizers. in *70th Device Research Conference* 211–212 (IEEE, 2012). doi:10.1109/DRC.2012.6256944.
7. Zhao, H. *et al.* Atomically-thin Femtojoule Filamentary Memristor. Advanced Materials 29.47 (2017): 1703232.
8. Clerc, S. *et al.* A 0.32V, 55fJ per bit access energy, CMOS 65nm bit-interleaved SRAM with radiation Soft Error tolerance. in *2012 IEEE International Conference on IC Design & Technology* 1–4 (IEEE, 2012). doi:10.1109/ICICDT.2012.6232860.
9. Wang, K. L., Alzate, J. G. & Khalili Amiri, P. Low-power non-volatile spintronic memory: STT-RAM and beyond. *J. Phys. Appl. Phys.* **46**, 074003 (2013).